\documentclass[preprint,preprintnumbers,amssymb,amsmath,9pt, superscriptaddress]{revtex4}[10pt]
\usepackage{graphicx}
\usepackage{dcolumn}
\usepackage{bm}
\begin{document} 
\input epsf.tex
\newcommand{\beq}{\begin{eqnarray}}
\newcommand{\eeq}{\end{eqnarray}}
\newcommand{\nn}{\nonumber}
\def\ltap{\ \raise.3ex\hbox{$<$\kern-.75em\lower1ex\hbox{$\sim$}}\ }
\def\gtap{\ \raise.3ex\hbox{$>$\kern-.75em\lower1ex\hbox{$\sim$}}\ }
\def\CO{{\cal O}}
\def\CL{{\cal L}}
\def\CM{{\cal M}}
\def\tr{{\rm\ Tr}}
\def\CO{{\cal O}}
\def\CL{{\cal L}}
\def\CN{{\cal N}}
\def\mpl{M_{\rm Pl}}
\newcommand{\bel}[1]{\be\label{#1}}
\def\al{\alpha}
\def\bt{\beta}
\def\eps{\epsilon}
\def\eg{{\it e.g.}}
\def\ie{{\it i.e.}}
\def\mn{{\mu\nu}}
\newcommand{\rep}[1]{{\bf #1}}
\def\be{\begin{equation}}
\def\ee{\end{equation}}
\def\bea{\begin{eqnarray}}
\def\eea{\end{eqnarray}}
\newcommand{\eref}[1]{(\ref{#1})}
\newcommand{\Eref}[1]{Eq.~(\ref{#1})}
\newcommand{\gsim}{ \mathop{}_{\textstyle \sim}^{\textstyle >} }
\newcommand{\lsim}{ \mathop{}_{\textstyle \sim}^{\textstyle <} }
\newcommand{\vev}[1]{ \left\langle {#1} \right\rangle }
\newcommand{\bra}[1]{ \langle {#1} | }
\newcommand{\ket}[1]{ | {#1} \rangle }
\newcommand{\eV}{{\rm eV}}
\newcommand{\ev}{{\rm eV}}
\newcommand{\kev}{{\rm keV}}
\newcommand{\Mev}{{\rm MeV}}
\newcommand{\gev}{{\rm GeV}}
\newcommand{\tev}{{\rm TeV}}
\newcommand{\mev}{{\rm MeV}}
\newcommand{\meV}{{\rm meV}}
\newcommand{\mnu}{\ensuremath{m_\nu}}
\newcommand{\nnu}{\ensuremath{n_\nu}}
\newcommand{\mlr}{\ensuremath{m_{lr}}}
\newcommand{\acc}{\ensuremath{{\cal A}}}
\newcommand{\mav}{MaVaNs}
\newcommand{\kt}{\ensuremath{k_{tach}}}
\newcommand{\zt}{\ensuremath{z_{tach}}}
\newcommand{\bzt}{\ensuremath{{\bar z}_{tach}}}
\title{Late Forming Dark Matter in Theories of Neutrino Dark Energy}
\author{Subinoy Das}
\affiliation{Center for Cosmology and Particle Physics,\\
  Department of Physics, New York University,\\
New York, NY 10003, USA}
\author{Neal Weiner}
\affiliation{Center for Cosmology and Particle Physics,\\
  Department of Physics, New York University,\\
New York, NY 10003, USA}
\preprint{ }
\date{\today}
\begin{abstract}
We study the possibility of Late Forming Dark Matter (LFDM), where a scalar field, previously trapped in a metastable state by thermal or finite density effects, begins to oscillate near the era matter-radiation equality about its true minimum. Such a theory is motivated generally if the dark energy is of a similar form, but has not yet made the transition to dark matter, and, in particular, arises automatically in recently considered theories of neutrino dark energy. If such a field comprises the present dark matter, the matter power spectrum typically shows a sharp break at small, presently nonlinear scales, below which power is highly suppressed and previously contained acoustic oscillations. If, instead, such a field forms a subdominant component of the total dark matter, such acoustic oscillations may imprint themselves in the linear regime.
\end{abstract}
\maketitle
\section{Introduction}
The increasingly significant evidence for the dark universe has established a strong paradigm in cosmology, in which the dynamics of the universe at the largest scales are governed by two components of energy which, up to this point, have only been observed by their gravitational consequences \cite{Colless:2001gk,Tegmark:2003uf,Riess:1998cb,Perlmutter:1998np,Spergel:2006hy}. These two, dark matter and dark energy, appear to behave in fundamentally different ways, with dark matter clustering into galaxies and diluting as the universe expands, while dark energy appears to remain smooth and dilutes either slowly or not at all, with equation of state near $w=-1$.

In spite of this, there is great effort to explore whether or not these substances might somehow be related. The strongest motivation for this is the similarity of the energy densities of $\rho_{DM}$ and $\rho_{DE}$ at the present epoch. Such attempts to connect these substances inevitably must confront the above differences, and attempts to unify them into one fluid often leads to dramatic observational consequences (see, for example \cite{Sandvik:2002jz}).

There is a slightly more restrained approach, however. Rather than viewing these substances as necessarily the same fluid, we might instead view them of a similar type. That is, dark matter may be a substance which, at some time in the past behaved as dark energy, and dark energy may, in the future, behave as dark matter. The fact that physics in the standard model has a generational structure, with repeated fields at different mass scales especially motivates such duplication. In particular, in theories where the dark energy is connected to a new neutrino force as recently explored in \cite{Fardon:2003eh,Peccei:2004sz,Fardon:2005wc}, such generational structure is expected in the dark energy sector.

Such a consideration immediately raises the question: for how long must dark matter have behaved as dark matter? Certainly, at least since matter radiation equality dark matter has been clustering and diluting more or less as $a^{-3}$. However, even at eras earlier than this, the clustering behavior of the dark matter can be observed in the power spectrum, at least to scales of $0.1h^{-1}\ Mpc$, where the matter power spectrum becomes non-linear.

It is quite natural to consider a scalar field which at some point in the history of the universe transitions to a dark matter state. Chaotic inflation \cite{Linde:1981mu,Albrecht:1982wi}, for instance, ends when the slow roll condition ends, and, for a suitable potential, begins to evolve as dark matter. A very familiar example of such dark matter is the axion \cite{Abbott:1982af,Dine:1982ah,Preskill:1982cy}, which acquires a (relatively) large mass after the QCD phase transition, at which point it begins to behave as dark matter. A conversion to dark matter is the natural final state of numerous quintessence theories \cite{Peebles:1987ek,Frieman:1995pm,Chacko:2004ky}

Our focus here will be on a transition that occurs much later in the universe, in order to make connections to theories of dark energy. In fact, we shall see that this transition naturally occurs near the era of matter-radiation equality. With such a late-time transition, effects on the CDM power spectrum are possible.
This ``Late Forming Dark Matter'' (LFDM), arises simply from a scalar field coupled to a thermal bath, initially sitting in a metastable state, behaving like a cosmological constant. When the thermal bath dilutes, the scalar transitions near matter-radiation equality (MRE) to dark matter, yielding interesting observable consequences.

The layout of this paper is as follows: in section \ref{sec:lfdm}, we lay out the basic structure of a general theory. In section \ref{sec:pheno} we will explore the effects of such a scenario on the power spectrum of dark matter. In section \ref{sec:neutrinos} we will see how this sort of dark matter naturally arises in theories of neutrino dark energy. In section \ref{sec:experiments} we will review what experimental studies constrain this scenario, and may test it in the future. Finally, in section \ref{sec:conclusions} we conclude.

\section{Late Forming Dark Matter}
\label{sec:lfdm}

Let us consider a single scalar field $\phi$ coupled to some other relativistic particle $\psi$ which is in thermal equilibrium. For simplicity, we will assume that $\phi$ is at zero temperature (i.e., its couplings to $\psi$ are sufficiently small that it is not thermalized). At zero temperature for $\psi$, we assume a potential of the form
\be
V(\phi) = V_0-\frac{m^2}{2} \phi^2 - \epsilon \phi^3 + \frac{\lambda}{4} \phi^4,
\ee
where $V_0$ is a constant which sets the true cosmological constant to zero. We assume the presence of the thermal $\psi$ contributes a term to the potential
\be
\delta V = D T^2 \phi^2
\ee
where $D$ is a coefficient determined by the spin, coupling, and number of degrees of freedom in $\psi$. 

The behavior of this system is simple to understand. At high temperature, there is a thermal mass for $\phi$ which will confine it to the origin. At
\be
T = \sqrt{\frac{\lambda m^2 + 2 \epsilon^2}{2 D \lambda}}
\ee
a new minimum appears at energy lower than at $\phi=0$. However, because of the thermal mass, $\phi$ remains trapped at the origin.

At a temperature
\be
T_{tach}=\frac{m}{\sqrt{2 D}}
\ee
$\phi$ becomes tachyonic about the origin, and will begin to oscillate about what then evolves into its true minimum. These oscillations then behave as dark matter. Note that the energy in the dark matter is set by the depth of the global minimum  relative to $\phi=0$ at $T_{tach}$, in this case $O(\epsilon^4/\lambda^3)$. If all the dimensionful parameters are of the same order (i.e., $\epsilon \sim m$), then the temperature at which dark matter is formed is soon followed by matter-radiation equality. Such correlation leaves a strong imprint on the power spectrum which will discuss in section \ref{sec:pheno}.

The above gives an extremely simple example of a model in which for most of the history of the universe, $\phi$ acted as a cosmological constant and only at very late times does $\phi$ begin to behave as conventional dark matter. 
Such a form of dark matter is very natural when similar structures explain the existence of dark energy, for instance in neutrino theories of dark energy.

\section{Cosmological Consequences}
\label{sec:pheno}
Unlike weak-scale dark matter, which necessitates some interactions with ordinary matter which may be tested at underground experiments, and unlike axions, which require a coupling to photons giving again an experimental test, LFDM theories need not have strong couplings to standard model fields. Even within theories of neutrino dark energy, where LFDM is motivated, direct tests are difficult, if not impossible.

The best hope of detection for such a scenario is cosmological. Because we expect $z_{tach}$ naturally to lie near $z_{MRE}$, we expect deviations in the power spectrum of DM at small $(k \gsim h Mpc^{-1})$ scales.
In this section we will discuss the signatures of LFDM and the predictions it makes for cosmological experiments.

In general, for our scenario, effects on the CMB are negligible. We will return to this issue later. As LFDM behaves as ordinary CDM after $z_{tach}$, we should not expect visible consequences on scales $k<k_{tach}$, where $k_{tach}$ is the scale of the horizon at $z_{tach}$.

\subsection{Power Spectra}

Let us consider the power spectrum for dark matter near \zt. Since this is when CDM is formed, after this point we can evolve it quite simply. The relevant quantity for the local density of dark matter is the redshift when it formed. Since all dark matter forms with the same initial energy density, regions where it forms earlier will have diluted more at later times, and regions where it forms later will have diluted less.

Dark matter forms at $\zt = \bzt + \delta \zt (x)$. By definition $\zt$ is the redshift when $T(\zt,x) = T_{ tach}$. We can reexpress the local temperature as 
\bea
T(\bzt + \delta z(x)) &=& \bar T(\bzt+\delta z)+\delta T(\bzt + \delta z,x)\\&=&(\bar T(\bzt)+\delta T(\bzt ,x))\frac{(1+\bzt)}{(1+\bzt + \delta z)}
.
\eea
In the last equality is clearly true only for regions over which sound waves cannot propagate between $\bzt$  and $\bzt  +\delta \zt$. Since this will be at scales of order $10^5$ smaller than the horizon size, we can neglect it for our purposes. By definition, $T_{tach}=T(\bzt+\delta \zt,x)=
{\bar T} (\bzt)$, and thus we can easily find that
\be 
\delta T(\bzt,x)/\bar T(\bzt) = \delta \zt/(1+\bzt)
\ee
Similarly, $\rho(z,x)/\bar \rho(z) =(1+\bzt)^3/(1+\bzt + \delta z(x))^3$, from which we can find
\be
\delta \rho(\bzt,x)/\bar \rho (\bzt) = 3 \delta \zt(x)/(1+ \bzt) = 3 \delta T(\bzt, x)/T(\bzt).
\ee Thus, at $z=\zt$ the CDM power spectrum is proportional to the $\psi$ temperature power spectrum at \zt. From this point, the density perturbations will grow as CDM, so determining the power spectrum of CDM today is tantamount to determining the $\psi$-temperature power spectrum at \zt.

We will ultimately want to identify $\psi$ with a more conventional particle-physics candidate, and, in particular, the neutrino.
  In general, the neutrino is highly relativistic at the time of its decoupling, 
   after which  it free-streams until it becomes non relativistic, yielding a suppression of its power at scales 
   $k>  k_{fs}=0.018 \Omega^{1/2} (\frac{m_{\nu}}{eV})^{1/2}$$Mpc^{-1} $.   However, in models of neutrino dark energy, there are additional neutrino interactions, and these may serve to keep the neutrino tightly coupled until \zt. If this is the case, this should be imprinted on the CDM power spectrum. Similar studies have been performed for scenarios where the neutrino was significantly heavier, and such strong interactions were proposed in order to retain neutrinos as dark matter
\cite{Raffelt:1987ah}. 
More recently, the implications of such neutrino interactions for cosmology have been studied \cite{Beacom:2004yd,Bell:2005dr,Cirelli:2006kt,Sawyer:2006ju}.

 \subsection{Calculation of power spectra for LFDM}

We will consider LFDM with both an interacting and a non-interacting coupled bath. As described above, we will compute the power spectrum of the relativistic fluid, and match that to the initial power spectrum of the CDM at $z=\zt$. The non-interacting case is straightforward. The interacting case can be got by considering earlier studies of the evolution of density perturbations for interacting neutrinos \cite{Atrio-Barandela:1996ur,Hannestad:2004qu}, where the the interaction makes different components behave as a single tightly coupled fluid. Under this assumption, the shear or anisotropic stress in the perturbation is negligible. The evolution is characterized by density and velocity perturbations only, and we can truncate all the higher order moments. The evolution of density and velocity perturbations is given by \cite{Ma:1995ey}
\be
\dot{\delta}=-(1+w) (\theta +\dot{h}/2)-3\frac{\dot{a}}{a} (c_{s}^{2}-w)\delta
\ee
\be
\dot{\theta}=-\frac{\dot{a}}{a} (1-3w) \theta -\frac{\dot{w}}{1+w} \theta + \frac{c_{s}^{2}}{1+w}k^{2} \delta
\ee
We are interested in the case where the thermal bath is made of essentially massless particles, so the equation of state and sound speed are given by $w=1/3=c_{s}^{2}$. To get the amplitude of the perturbation at any redshift, the above two equations need to be solved with the background equations of motion for the metric perturbations. We have modified the publicly available CAMB and CMBFAST to solve and get the power spectra at $z_{tach}$.

\begin{figure*}[t]
a)
\includegraphics[width=120mm]{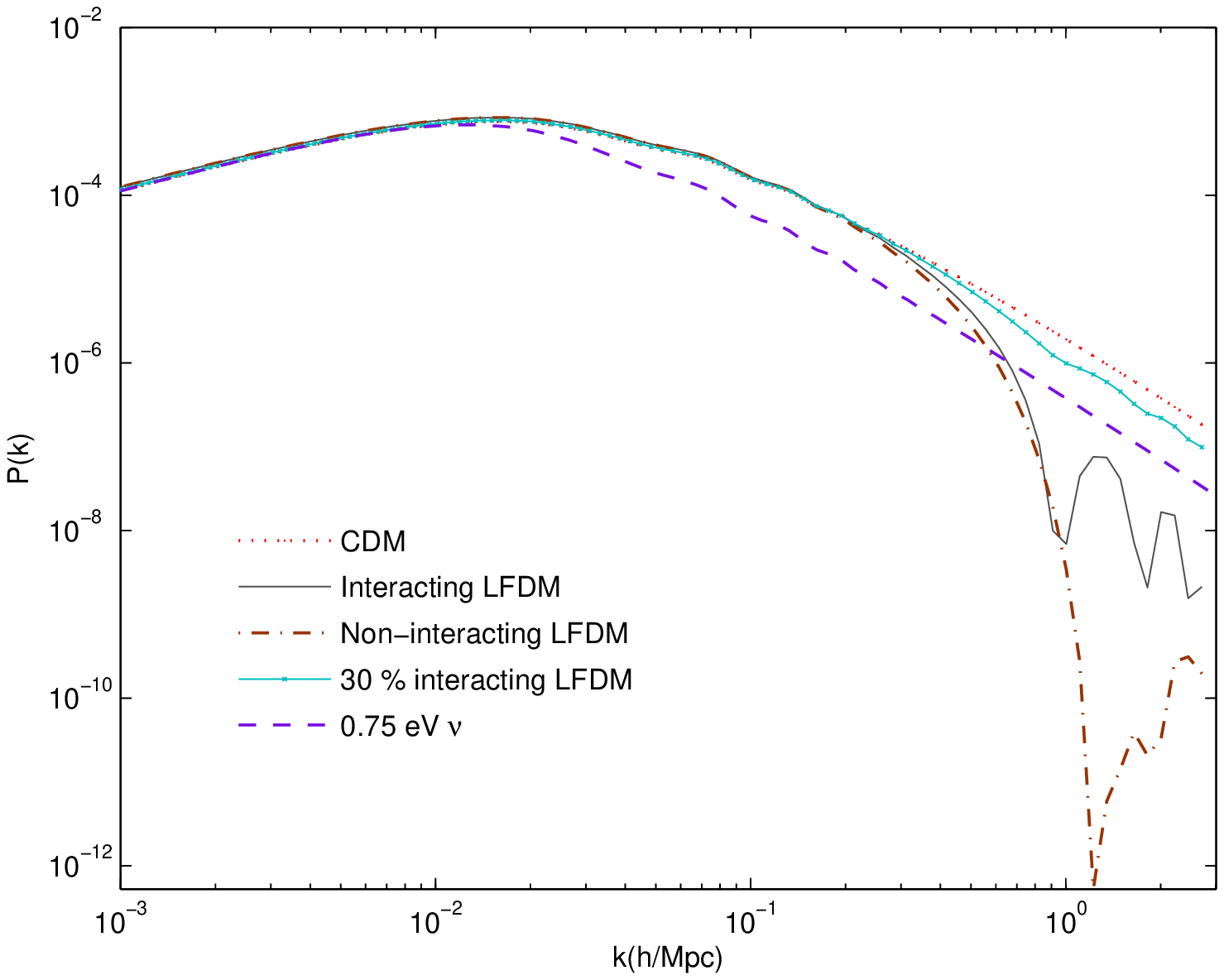}
\vskip 0.01in 
b) \includegraphics[width=120mm]{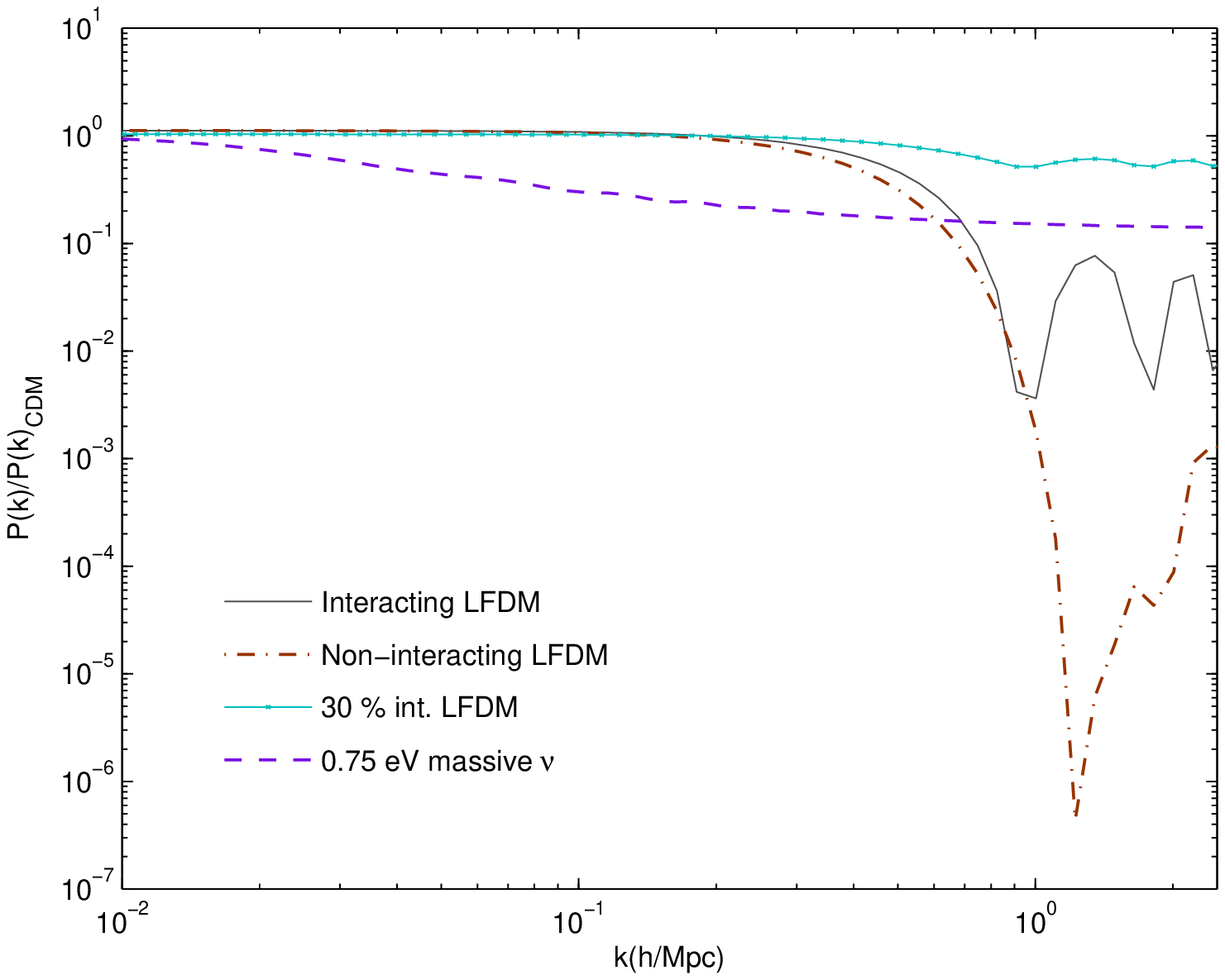}
\caption{Power spectra (a), and power compared to CDM (b) for CDM, LFDM (with different fractions, interacting and free-streaming) and a 0.75 eV freestreaming neutrino for $z_{tach}=50,000$
}
\label{fig:LFDMpower}
\end{figure*}

After \zt, LFDM follows the same evolution equation as CDM, and it is straightforward to grow the perturbations to today. 
 We are principally interested in situations where LFDM makes up all or nearly all or the dark matter, but we can also consider situations where it is only some fraction. As we see in figure \ref{fig:LFDMpower}, there is a suppression of power at small scales, and the possibility of acoustic oscillations imprinted on the power spectrum. 
For comparison, we also include the power spectrum for $\Lambda$CDM with a $0.75$ eV massive neutrino, near the experimental limit
\cite{Seljak:2004xh,MacTavish:2005yk,Hannestad:2004bu}. Though both LFDM and a massive neutrino give  suppression in power, there is a distinct difference in power spectra between the two. The suppression of power for a massive neutrino turns on much more gradually than the abrupt suppression for LFDM.

 As we make \zt\ smaller (larger), we move the break to larger (smaller) scales. At scales much smaller than \kt\ we would expect the acoustic oscillations to be damped out (which is not captured by our tightly coupled approximation). If LFDM is merely a fraction of the dark matter, the observability of such oscillations would depend on how much LFDM existed. If LFDM is all or nearly all of the dark matter, the oscillations are already severely constrained, and must lie in the non-linear regime \footnote{Though in different context,  \cite{Mangano:2006mp} has found similar oscillations for power spectra of a neutrino interacting with dark matter.}.

  It is important to point out here, though we get a large suppression beyond $k\approx 0.01 h Mpc^{-1}$ we cannot 
compare it directly to the linear power spectra of standard $\Lambda$CDM cosmology in this regime as the the non-linear effects in structure formations \cite{nl1,seljak} become very important for $k \gsim 0.15 h Mpc^{-1}$. We return to this issue in section \ref{sec:experiments}. Only if LFDM forms later in time ($z_{tach} \ll 15000 $), does the power get suppressed in the linear regime. In this case a rigorous statistical analysis would be needed to place legitimate constraints on this scenario, which is beyond the scope of this paper. 

\section{Models of LFDM in theories of Neutrino Dark Energy}
 \label{sec:neutrinos}
The idea of LFDM is appealing, largely because it offers to make a connection to theories of dark energy. If the dark energy is associated with a scalar field trapped at a false minimum in its potential due to thermal effects, then, quite likely, ``copies'' of such physics may have existed earlier. If so, the energy stored there would now behave as dark matter. 

Remarkably, there is already a class of models that fit these criteria, specifically the recently discussed ``hybrid'' models of neutrino dark energy \cite{Fardon:2005wc}.
There has been a long motivation to make a connection between neutrino masses and dark energy  \cite{Hill:1988vm,Hung:2000yg,Gu:2003er,Fardon:2003eh,Peccei:2004sz}.  In these most recent models, the generational structure of the neutrinos is copied in the dark energy sector. The finite density of relic neutrinos modifies the potential and stabilizes a scalar field at a false minimum. These hybrid models arise naturally when MaVaN models are promoted to a supersymmetric theory (see \cite{Takahashi:2005kw,Takahashi:2006ha,Takahashi:2006be} for other supersymmetric extensions).

The natural extension to LFDM comes in these supersymmetric theories. We refer readers to \cite{Fardon:2005wc} for details, and only briefly summarize here. Because there are three neutrinos, these theories contain three singlet neutrinos $N_i$. Each one of these is associated by supersymmetry with a scalar field. Arguments related to naturalness suggest the the lightest of the three neutrinos is associated with the dark energy today. Energy previously trapped in the other scalar neutrinos would appear as dark matter today, and it is this that we consider.

We shall now present a simple model of LFDM within the context of neutrino dark energy theories. It is not intended to be representative of all such models, but merely a simple example of one with the relevant phenomenology.

Consider the fermion fields $\psi_{2,3}$, and scalars $n_{2,3}$, with a Lagrangian
\be
{\cal L}=\lambda n_2 \psi_3^2+ 2 \lambda \psi_2 \psi_3+m_3 \psi_3 \nu_3+m_2 \psi_2 \nu_2+V_{susy}+V_{soft}+V_{\epsilon}
\ee
where 
\be
V_{susy} = 4 \lambda^2 |n_2|^2 |n_3|^2 + \lambda^2 |n_3|^4,
\ee
\be
V_{soft} = \tilde m_2^2 |n_2|^2-\tilde m_3^2 |n_3|^2+ (\tilde a_3 n_3^3+h.c.)
\ee
and
\be
V_{\epsilon} =   4 \lambda \epsilon (n_3^*  n_2^3 + n_3^3 n_2^*+h.c.) + \epsilon^2 (|n_2|^4 + 4 |n_3|^2 |n_2|^2).
\ee
Such a Lagrangian can easily be constructed supersymmetrically with soft terms of their natural size. The terms in $V_\epsilon$ are included in order to generate a Majorana mass for the neutrino in the vacuum.
We also expect couplings to the ``acceleron'' (again, see \cite{Fardon:2003eh,Fardon:2005wc}), which is directly tied to the stability of dark energy today. Both these couplings as well as $V_\epsilon$ do not influence our discussion here. It has been demonstrated that the vevs of such fields do not spoil the success of the dark energy theory in these hybrid models \cite{Spitzer:2006hm}.
 
The natural size of each soft term is of the order of the associated Dirac mass (i.e., $\tilde a_3 \sim \tilde m_3 \sim m_3$ which is expected to be of order 0.1 eV), assuming the dark energy sector has no approximate R-symmetry.
 
If $n_2$ has a large expectation value, it generates a Majorana mass for $\psi_3$ of order $m_3^2/{\lambda n_2}$. The presence of the relic neutrinos affects the dynamics of $n_2$, in particular by driving it to larger values. 
 Assuming that the relic neutrinos are in the light mass eigenstate (see \cite{bbn}), the relic neutrinos contribute a term to the effective potential for $n_2$,
 \be
 V_{eff} = \frac{T^2 m_3^4}{24 \lambda^2 n_2^2}
 \ee
 which is minimized for large $n_2$, competing with the $n_2$ mass term which is minimized at $n_2=0$. The competition drives an expectation value $\vev{\lambda n_2} \sim m_3 \sqrt{\lambda T/m_2}$. (We should note all temperatures here refer to neutrino temperature, which are slightly lower than the CMB temperature.)  The non-zero value of $n_2$ creates a positive value for the mass squared of $n_3$, stabilizing it in the false vacuum with an effective cosmological constant. Such a model is analogous to hybrid inflation models, with $n_2$ playing the role of the slow-roll field, and $n_3$ playing the role of the waterfall field.
 
 The temperature where $n_3$ becomes tachyonic is $T_{\rm tach}=\sqrt{3/2}  m_2 \tilde m_3^2/\lambda m_3^2$, and the energy converted to dark matter at that time is $\rho_{LFDM} \sim 10^{-3} \tilde a_3^4/\lambda^6$. The time of matter radiation equality is $T_{MRE} = 3 \sqrt{3/2} \tilde a_3^4 m_3^6/64 \lambda^3 m_2^3 \tilde m_3^6 $. Because of the high powers of parameters, each uncertain by factors of order one, there is a high uncertainty in $T_{MRE}$. Simply varying the mass parameters in the ranges $10^{-1.5} eV <\tilde m_3, \tilde a_3,m_3 < 10^{-.5}$, $10^{-2} eV < m_2 <  10^{-1} eV$ and the parameter $10^{-2} <\lambda < 1$, we find $10^{-3} eV \lsim T_{MRE} \lsim 10^7 eV$. Similarly, we find $10{-1}\lsim T_{MRE}/T_{DMDE} \lsim 10^{13}$. Hence, the solution to the coincidence problem is present in that such a crossing should occur relatively soon after matter-radiation equality. However, the precise value is clearly uncertain, so the success is limited.
 
 A more precise statement of the success of the solution to the coincidence problem is that {\em if} this is the explanation of the present coincidence (by relating the energy densities to neutrino masses), then a break should exist in the power spectrum. Given that we can set $\lambda$ by fixing $T_{MRE}$, we can more precisely determine $T_{tach}$, even with the uncertainties of parameters. Thus, using the same ranges above, and requiring $\lambda<1$, one finds that $1 \ev \lsim T_{tach} \lsim 10^3 \ev$ and thus $2\times 10^{-2} h Mpc^{-1} \lsim k_{tach} \lsim 20 h Mpc^{-1}$. Such limits are certainly model dependent, but clearly there is a strong expectation of a break in the power spectrum in the observable range.

\section{Experiments}
\label{sec:experiments} 
A great deal of data already would constrain such a scenario. For instance, 
one immediate concern would be from the CMB.
In general, neutrinos are not freestreaming at recombination, which affects the gravitational potential well which boosts the first peak of the CMB. Such constraints have been considered \cite{Hannestad:2005ex,Bell:2005dr}, but one interacting neutrino seems acceptable. 
One also must consider ($1.6 \leq N_{eff}^{\nu} \leq 7.1 $), the constraint on total number of freestreaming neutrinos during decoupling because 
having extra radiation degrees of freedom could delay the matter radiation equality resulting in the early ISW effect\cite{Crotty:2003th,Hannestad:2003xv,Pierpaoli,Hannestad:2006mi}. 
Structure formation is where LFDM is most likely to be tested. Many experiments such as 2dF Galaxy Redshift survey \cite{Colless:2001gk,Percival:2001hw}, Sloan Digital Sky Survey (SDSS) \cite{SDSS}, Ly-$\alpha$ forest \cite{Viel:2004bf,Viel:2004np,Seljak:2006bg,Viel:2005ha,McDonald:2004eu}, and weak gravitational lensing \cite{Refregier:2003xe} have measured the matter power spectrum over a wide range of scales. Though these experiments are in good agreement with the $\Lambda$CDM model, small scales remain an open question, with a possible modifications seen in Lyman-$\alpha$ systems \cite{seljak}, as well as some studies of dwarf galaxies \cite{Gilmore:2006iy}.

The studies most promising to test this scenario in the future would include Ly-$\alpha$ data, but one still needs non-linear simulations to extract the linear power spectra information on these length scales. Future weak lensing experiments \cite{Abazajian:2002ck} will measure the power at higher $z$ when the relevant scales would be more linear.  
Other experiments like 21 cm tomography \cite{Loeb:2003ya} will also measure power in very small scales (sub-Mpc) and may find signatures of LFDM. As discussed before, to compare LFDM power spectra with experiments in this range we need detailed N-body
 simulation which includes the non-linear hydrodynamical effects of gravitational clustering.

 \section{Conclusions}
\label{sec:conclusions}
We have considered the scenario of ``Late Forming Dark Matter'' (LFDM) in which a scalar field converts the energy of a metastable point to dark matter at times late in the history of the universe, near the era of matter-radiation equality. Such effects arise when the potential of the scalar field is strongly effected by finite temperature effects from some additional thermal species. These theories arise naturally in hybrid models of neutrino dark energy, in which new scalar fields arise in association with neutrinos.

The power spectrum of such theories naturally has a sharp near the scale of the horizion at matter-radiation equality, due to the streaming of the thermal species. The presence strong scattering of these particles can modify the depth of the break, and the presence of acoustic oscillations.

Within the context of theories of neutrino dark energy, the scale of dark energy is controlled by the scale of neutrino masses, and, similarly, the amount of dark matter, and the redshift at which it forms, \zt, are also determined by the neutrino masses. In these simple theories, consistency requires a sharp break in the CDM power spectrum in the approximate range $10^2\ h Mpc^{-1} \gsim k_{tach} \gsim 10^{-3}\ h Mpc^{-1}$. Future studies at small scales, such as of Lyman-$\alpha$ systems, gravitational lensing or 21 cm absorption may be able to test these theories.

\section*{Acknowledgements} We thank Kris Sigurdson for his help in including interactions into CMBFAST and CAMB, as well as Roman Scoccimarro and Ann Nelson for reading a draft of the manuscript and providing useful comments.  This work was supported by NSF CAREER grant PHY-0449818 and by the DOE OJI program under grant DE-FG02-06ER41417. 

\

\bibliography{lfdm}
\bibliographystyle{apsrev}
\end{document}